\documentclass[%
 reprint,
superscriptaddress,
nofootinbib,
 amsmath,amssymb,
 aps,
]{revtex4-2}
\usepackage{subcaption}

\usepackage{graphicx}
\usepackage{dcolumn}
\usepackage{bm}

\usepackage{booktabs}
\usepackage{xcolor,colortbl}
\usepackage{hyperref}

\begin{document}

\preprint{APS/123-QED}

\title{Equation of State at High Baryon Densities from a Thermodynamically Informed Neural Network}


\author{Musfer Adzhymambetov}
\email{adzhymambetov@gmail.com}
\affiliation{%
Bogolyubov Institute for Theoretical Physics,
Metrolohichna  14b, 03143 Kyiv,  Ukraine 
}%


\date{\today}

\begin{abstract}

We present a four-dimensional equation of state for strongly interacting matter at finite temperature and conserved charge densities, constructed using a deep neural network. It is designed for direct use in hybrid models of relativistic heavy-ion collisions: it reproduces hadron resonance gas thermodynamics at typical particlization scales, is consistent with lattice QCD at low baryon chemical potential, and extrapolates into the high-density region inaccessible to either approach, which is precisely the regime targeted by RHIC BES, FAIR, HADES, and CBM.  Thermodynamic consistency throughout the full phase space is enforced via a physics-informed loss function. We demonstrate the developed equation of state by implementing it at zero net strangeness and fixed electric-to-baryon charge ratio within the integrated hydrokinetic model.
\end{abstract}


\maketitle

\section{Introduction}
Understanding the thermodynamic properties of strongly interacting matter remains one of the central open problems in nuclear and particle physics. Quantum Chromodynamics (QCD) provides the underlying theoretical framework, but its non-perturbative nature makes direct calculations extremely challenging except in limited regimes. At zero baryon density, lattice QCD simulations yield high-precision results \cite{HotQCD:2014kol, Borsanyi:2010cj} for thermodynamic quantities such as pressure, energy density, and entropy density. In the confined phase, the hadron resonance gas  (HRG) model reproduces these results up to temperatures near the crossover \cite{Huovinen:2009yb, Vovchenko:2014pka, Andronic:2012ut}. At finite baryon density, however, lattice calculations are obstructed by the sign problem \cite{deForcrand:2009zkb}, blocking direct access to the region of the phase diagram probed by ongoing and future heavy-ion experiments at the RHIC Beam Energy Scan \cite{STAR:2010vob}, HADES \cite{HADES:2009aat}, and CBM \cite{CBM:2016kpk}. In these programs, the baryon chemical potential at chemical freeze-out can reach around 800 MeV \cite{Cleymans:2005xv, Andronic:2017pug}, and the system becomes sensitive to a possible phase transition and critical endpoint \cite{Stephanov:1998dy, Bzdak:2019pkr}.

In this regime, theoretical constructions rely on phenomenological models. The MIT Bag Model~\cite{Chodos:1974je} was among the first, describing deconfined matter as a gas of non-interacting quarks and gluons confined by a constant vacuum pressure and providing an early quantitative picture of a first-order transition to the quark-gluon plasma. More advanced frameworks followed: chiral mean-field models~\cite{Papazoglou:1998vr, Motornenko:2019arp} incorporate chiral symmetry restoration and parity doubling in the hadronic sector, while quasiparticle models \cite{Peshier:1999ww, Schneider:2001nf} describe the quark-gluon plasma through effective degrees of freedom with medium-dependent masses. Both classes are calibrated to lattice QCD data at zero baryon density and then extrapolated to finite density, yielding controlled but model-dependent extensions of the equation of state (EoS).

Non-parametric methods have been developed to reduce this model dependence. Interpolation schemes based on hyperbolic tangent functions, cubic splines, or polynomial fits \cite{Huovinen:2009yb, Albright:2014gva, Monnai:2019hkn} connect HRG results at low temperatures to lattice data at high temperatures, while Gaussian process regression~\cite{Gong:2024lhq} reconstructs the EoS without assuming a fixed functional form. These approaches are efficient and numerically stable, but rely on externally imposed matching prescriptions and choices for the transition region. Bayesian inference has been used to constrain the remaining freedom: in~\cite{OmanaKuttan:2022aml}, a density-dependent EoS is constrained against heavy-ion observables; in~\cite{Liu:2023rfi}, against lattice data at zero baryon density; and in~\cite{Ayriyan:2024zfw}, against pulsar and neutron star observations at high density.

Machine learning offers a complementary non-parametric approach. Neural networks act as universal function approximators for thermodynamic relations without assuming a fixed analytic form, and automatic differentiation allows them to be embedded directly within physics solvers. Applications to neutron star matter have demonstrated the viability of this strategy~\cite{Soma:2022qnv, Morawski:2020izm}. A key limitation of standard networks, however, is poor extrapolation outside the training domain — a serious drawback for the QCD EoS, where data are sparse and confined to narrow regions of the phase diagram. Physics-informed neural networks (PINNs) address this by incorporating physical laws directly into the loss function \cite{raissi2017ph}, penalizing violations of the governing equations across the full domain. This approach has been applied to parameterize quasiparticle masses at zero~\cite{Li:2022ozl} and finite~\cite{Li:2025csc} baryon density.

A separate but related challenge concerns the EoS as an input to heavy-ion simulation frameworks. Modern hybrid models couple a macroscopic hydrodynamic stage to a microscopic transport model for the dilute hadronic phase \cite{Bass:2000ib, Petersen:2008dd}. The EoS is the closure relation for the hydrodynamic equations, controlling the expansion rate and the buildup of collective flow anisotropies. It also governs particlization: the local temperature and chemical potentials on the switching hypersurface are read from the EoS and used in the Cooper-Frye prescription \cite{Cooper:1974mv} to sample the initial hadron distribution for the transport stage. This procedure assumes an HRG EoS, so for a smooth transition the hydrodynamic EoS must be consistent with the HRG at particlization — otherwise discontinuities in thermodynamic quantities distort the subsequent evolution \cite{Huovinen:2012is}.

In this work, we develop a physics-informed neural network framework to construct a four-dimensional equation of state for strongly interacting matter. The model takes the temperature and the three conserved-charge chemical potentials as inputs,
\begin{equation}\label{eq:X}
    X = (T,\mu_B,\mu_Q,\mu_S)\, ,
\end{equation}
and returns the pressure as output. All remaining thermodynamic quantities are obtained through automatic differentiation.

The construction is designed to simultaneously satisfy the constraints outlined above: it reproduces lattice QCD results where available, matches the HRG EoS along and below the particlization hypersurface, and provides a continuous extrapolation toward higher baryon chemical potentials over the domain, which we defined as
\begin{align}
\label{eq:domain}
\mathcal{R}_X =
\begin{cases}
T     &\in [0.02,\,0.5]~\mathrm{GeV}, \\
\mu_B &\in [-0.1,\,1.0]~\mathrm{GeV}, \\
\mu_Q &\in [-0.04,\,0.04]~\mathrm{GeV}, \\
\mu_S &\in [-0.1,\,0.4]~\mathrm{GeV}.
\end{cases}
\end{align}

Thermodynamic consistency is enforced across the full domain through dedicated PINN loss terms, while no specific microscopic model or built-in phase transition is assumed. The resulting framework constitutes a proof-of-concept EoS aimed at hybrid-model applications at low collision energies, including RHIC BES, HADES, CBM, and AGS, where large baryon densities are reached, and consistency with the HRG EoS at particlization is essential. In addition, it may serve as a smooth background EoS for studies involving critical-point embeddings \cite{Kapusta:2021oco, Parotto:2018pwx}.

\section{Equation of State}
\subsection{Thermodynamic Framework}

The equilibrium thermodynamics of $(2+1)$-flavor QCD matter is governed by the grand canonical partition function $\mathcal{Z}(V,T,\mu_B,\mu_Q,\mu_S)$. In the thermodynamic limit, the pressure
\begin{equation}
    P(T,\mu_i) = \frac{T}{V} \ln \mathcal{Z}(V,T,\mu_i)
\end{equation}
serves as the fundamental thermodynamic potential, from which all bulk observables follow by differentiation. The entropy density and conserved charge densities ($i \in \{B,Q,S\}$) are
\begin{equation}\label{eq:sn}
    s = \left( \frac{\partial P}{\partial T} \right)_{\mu_i},
    \qquad
    n_i = \left( \frac{\partial P}{\partial \mu_i} \right)_{T},
\end{equation}
and the energy density is reconstructed via the Gibbs--Duhem relation,
\begin{equation}\label{eq:eps}
    \varepsilon = -P + Ts + \sum_{i} \mu_i n_i.
\end{equation}
This structure ensures thermodynamic consistency, since all observables derive from a single scalar potential $P(T,\mu_i)$.

\subsection{Lattice QCD Input}

At vanishing chemical potentials, QCD thermodynamics is determined from first-principles lattice simulations, providing high-precision results for bulk observables as functions of temperature. The extension to finite chemical potentials is obstructed by the fermion sign problem; however, thermodynamic quantities can be expanded in powers of $\mu_i/T$ around $\mu_i = 0$.

Following Ref.~\cite{Noronha-Hostler:2019ayj}, a Taylor expansion truncated at fourth order is employed:
\begin{equation}\label{eq:LQCD_P}
    \frac{P_{\text{LQCD}}}{T^4}
    =
    \sum_{i+j+k \leq 4}
    \frac{\chi^{BQS}_{ijk}}{i!\,j!\,k!}
    \left( \frac{\mu_B}{T} \right)^i
    \left( \frac{\mu_Q}{T} \right)^j
    \left( \frac{\mu_S}{T} \right)^k,
\end{equation}
where the generalized susceptibilities are
\begin{equation}
\chi^{BQS}_{ijk}(T)
=
\left.
\frac{\partial^{i+j+k}(P/T^4)}
{\partial(\mu_B/T)^i \partial(\mu_Q/T)^j \partial(\mu_S/T)^k}
\right|_{\mu_i=0}.
\end{equation}

These susceptibilities were computed on $48^3 \times 12$ lattices in the temperature range $135~\mathrm{MeV} < T < 220~\mathrm{MeV}$~\cite{Borsanyi:2018grb}. In this work, we use the parametrizations of Ref.~\cite{Noronha-Hostler:2019ayj}, where these 
results are extended outside the original temperature range by matching to the HRG model at low temperatures and to the Stefan--Boltzmann limit at high temperatures, yielding smooth and thermodynamically consistent expressions. The Taylor expansion 
remains reliable for $\mu_B/T \lesssim 2$--$2.5$, beyond which higher-order contributions become significant.

\subsection{Hadron Resonance Gas Model}

In the confined phase, QCD thermodynamics is well reproduced by the Hadron Resonance Gas model~\cite{Huovinen:2009yb, Vovchenko:2014pka}. The model rests on the resonance gas hypothesis: an interacting hadronic system with a sufficiently dense resonance spectrum can be mapped onto a non-interacting gas of stable hadronic states, with interactions effectively encoded through the inclusion of resonances \cite{Dashen:1969ep, Venugopalan:1992hy}. Within this 
framework, the total pressure is a sum over all hadronic species,
\begin{equation}
    P_{\mathrm{HRG}}(T,\mu_i) = \sum_{h} P_h(T,\mu_h)\, ,
\end{equation}
where the partial pressure of species $h$ is given by the standard grand-canonical 
expression,
\begin{equation}\label{eq:HRG_P}
    P_h = \pm \frac{g_h T}{2\pi^2}
    \int_0^\infty p^2\, dp\,
    \ln \left[ 1 \pm e^{-(E_h - \mu_h)/T} \right],
\end{equation}
with $g_h$ the degeneracy factor, $E_h(p)=\sqrt{p^2 + m_h^2}$ the on-shell 
single-particle energy, and upper (lower) signs corresponding to fermions (bosons). 
The chemical potential of each species is determined by its conserved quantum numbers,
\begin{equation}
    \mu_h = B_h \mu_B + Q_h \mu_Q + S_h \mu_S\, ,
\end{equation}
where $B_h$, $Q_h$, and $S_h$ denote the baryon number, electric charge, and 
strangeness of species $h$, respectively. The hadron list used in this work is generated by the \texttt{hadronSampler} code~\cite{Karpenko:hadronSampler} and is available 
at~\cite{Adzhymambetov:2025eos}.

\subsection{Hybrid Equation of State Construction}

Our goal is to construct a four-dimensional equation of state that (i) reproduces 
lattice QCD results at small $\mu_B/T$, (ii) matches the HRG EoS at low densities, 
and (iii) provides a controlled extrapolation up to $\mu_B \sim 1~\mathrm{GeV}$. 
Such an EoS is required for hybrid simulations of heavy-ion collisions at beam 
energies $\sqrt{s_{NN}} \sim 2$--$10~\mathrm{GeV}$, where $\mu_B/T$ can be large 
and direct lattice calculations are inapplicable.

In hybrid frameworks, the early dense stage is modeled by relativistic hydrodynamics 
under the assumption of local thermodynamic equilibrium. As the system expands and 
cools, it is converted to particles via the Cooper--Frye 
prescription~\cite{Cooper:1974mv}. The contribution from a fluid cell with 
hypersurface element $\Delta\Sigma_\mu$ to the distribution of hadron species $i$ is
\begin{equation}
E \frac{dN_i}{d^3p}
=
f_i(x,p)\, p^\mu\, \Delta\Sigma_\mu,
\end{equation}
where $f_i(x,p)$ is the phase-space distribution evaluated in the local rest 
frame of the fluid cell, usually taken as an equilibrium HRG distribution 
supplemented by viscous corrections. This hypersurface is typically an isosurface of a thermodynamic quantity, reconstructed numerically using the Cornelius algorithm~\cite{Huovinen:2012is, Molnar:2014fva}. Common choices include constant temperature, energy density, or particle 
density~\cite{Anchishkin:2022hnj}; criteria based on the Knudsen number or 
mean free path have also been proposed~\cite{Ahmad:2016ods, Adzhymambetov:2024zzz, Goes-Hirayama:2025nls}. The Cooper--Frye formula ensures conservation of all charge fluxes though $\Delta\Sigma_{\text{sw}}$, though it can produce negative contributions from space-like surface 
elements~\cite{Oliinychenko:2014tqa, Huovinen:2012is, Sinyukov:2002if};  we do not address this issue here.

A key requirement is that the hydrodynamic EoS and the HRG EoS used in the 
transport stage agree on $\Delta\Sigma_{\text{sw}}$; otherwise, discontinuities arise in intensive quantities such as temperature and chemical potentials and distort the subsequent evolution. The construction proposed here directly addresses this requirement.

\begin{figure}
    \centering
    \includegraphics[width=1.0\linewidth]{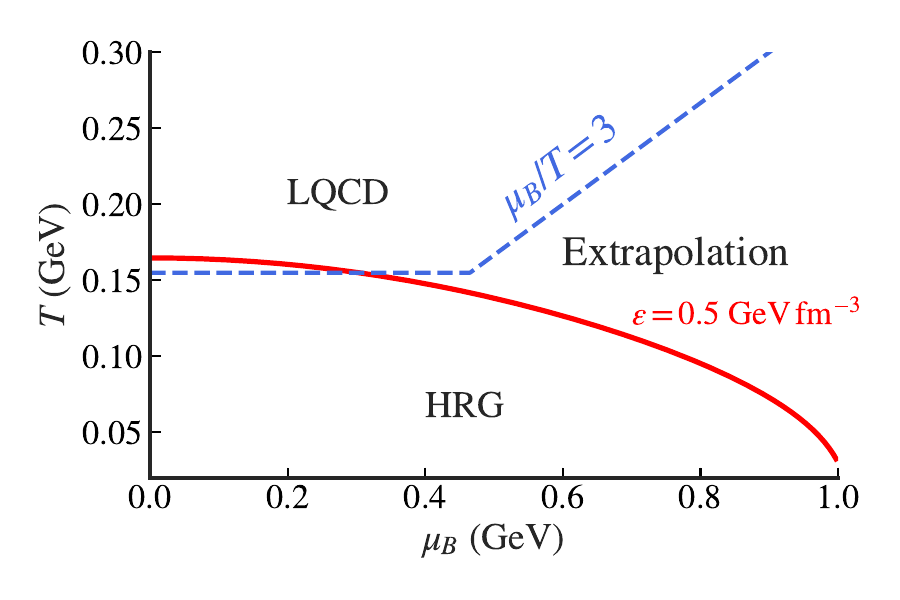}
    \caption{Schematic of the hybrid EoS construction in the $T$--$\mu_B$ plane. The LQCD region ($\mu_B/T \leq 3$), the HRG region (bounded by constant energy density $\varepsilon = 0.5~\mathrm{GeV/fm^3}$), and the extrapolation region are indicated.}
    \label{fig:diagram}
\end{figure}

The overall structure of the construction in the $T$--$\mu_B$ plane (at 
$\mu_Q = \mu_S = 0$) is summarized in Fig.~\ref{fig:diagram}. The EoS 
incorporates LQCD results for $\mu_B/T \leq 3$ and reproduces the HRG below 
$\varepsilon = 0.5~\mathrm{GeV/fm^3}$. Beyond these regions, it is extrapolated over the full domain $\mathcal{R}_X$ defined in Eq.~\eqref{eq:domain}.

\section{Neural Network Implementation}

\subsection{Network Architecture}

The equation of state is fully determined by the pressure $P(X)$, where 
$X = (T, \mu_B, \mu_Q, \mu_S)$. The neural network approximates a rescaled, 
preprocessed representation $Y$ of the pressure,
\begin{equation}\label{eq:ftheta}
f_{\theta}: X \mapsto Y,
\end{equation}
where $\theta$ denotes the trainable weights and biases. The explicit form of $Y$ 
is described in the next subsection. All other thermodynamic observables are 
obtained from $P_\theta(X)$ by automatic differentiation, as in 
Eqs.~(\ref{eq:sn}) and~(\ref{eq:eps}).

In the present work, we do not include a phase transition, and accordingly require 
$f_\theta$ to be $C^\infty$ smooth. The network is implemented in 
PyTorch~\cite{Paszke:2019fta} as a residual network (ResNet)~\cite{he2015} with analytic activation functions: an input layer with four neurons corresponding to $X$, followed by 16 residual blocks of 64 neurons each with $\tanh$ activations, and a 
final linear output layer producing $Y$. Thermodynamic observables are obtained by automatic differentiation through the network graph.

\subsection{Training Data and Preprocessing}

The training dataset 
\begin{equation}
    \mathcal{D} = \mathcal{D}_{\mathrm{LQCD}} \cup \mathcal{D}_{\mathrm{HRG}},
\end{equation}
is constructed from the lattice QCD and HRG equations of state defined in Eqs.~(\ref{eq:LQCD_P}) and (\ref{eq:HRG_P}). Each point contains $X$ together with observables $\{P, \varepsilon, s, n_B, n_Q, n_S\}$. Sampling is performed with domain restrictions to ensure physical relevance. For $\mathcal{D}_{\mathrm{LQCD}}$, uniform sampling is used over

\begin{equation}
\begin{aligned}
0.155 \le T \le 0.55~\mathrm{GeV}, \qquad
&|\mu_B|/T \le 3, \\
|\mu_Q|/T \le 0.2, \qquad
&|\mu_S|/T \le 0.6\, ,
\end{aligned}
\end{equation}

where the bounds for chemical potentials are motivated by the validity of the Taylor expansion.

For $\mathcal{D}_{\mathrm{HRG}}$, sampling is weighted as $|\mu_B|^3$ to enhance high-density coverage where the EoS is highly nonlinear. We again apply additional cuts motivated by the region that we can expect in heavy-ion collisions:

\begin{equation}
\begin{aligned}
T \ge 0.01~\mathrm{GeV}, \qquad
&\varepsilon > 0.01~\mathrm{GeV/fm^3}, \\
|\mu_Q| \le 0.06~\mathrm{GeV}, \qquad
&-0.1 \le \mu_S \le 0.4~\mathrm{GeV}.
\end{aligned}
\end{equation}

The range for $\mu_B$ is the same for both components
\begin{equation}
    -0.1 \le \mu_B \le 1.1~\mathrm{GeV}.
\end{equation}

After applying the cuts, we obtain the following dataset sizes
\begin{equation}\label{eq:Dsizes}
\begin{aligned}
|\mathcal{D}_{\mathrm{HRG}}| \approx 1.2 \times 10^{5}, &\qquad
|\mathcal{D}_{\mathrm{LQCD}}| \approx 0.8 \times 10^{5}, \\
N &\approx 2.0 \times 10^{5}.
\end{aligned}
\end{equation}

Since thermodynamic observables grow rapidly with temperature, we subtract the zero-density LQCD baseline:
\begin{align}
    \delta P &= P(T,\mu_i) - P_{\mathrm{LQCD}}(T,0), \\
    \delta s &= s(T,\mu_i) - s_{\mathrm{LQCD}}(T,0), \\
    \delta \varepsilon &= \varepsilon(T,\mu_i) - \varepsilon_{\mathrm{LQCD}}(T,0).
\end{align}

The network target is the renormalized pressure \begin{equation}
    Y = \frac{1}{\beta_P}\,\frac{\delta P}{(T+\delta T)^2},
\end{equation}
where $\delta T = 0.05~\mathrm{GeV}$ stabilizes low-temperature behavior and $\beta_P$ normalizes $Y$ to unit variance. The physical pressure is reconstructed as
\begin{equation}
    P_{\theta}(X) = \beta_P (T+\delta T)^2 f_{\theta}(X) + P_{\mathrm{LQCD}}(T,0).
\end{equation}

\subsection{Loss Function and Training}

The loss function combines data fidelity terms with physics-informed constraints. 
For each observable $\mathcal{O}_{i,j} \in \{\delta P,\, \delta\varepsilon,\, \delta s,\, 
n_B,\, n_Q,\, n_S\}_i$ we use a tolerance-modified mean squared error,
\begin{equation}\label{eq:tol_loss}
\mathcal{L}_{j} =
\frac{1}{N}\sum_{i=1}^{N}
\max\!\left[
0,\,
\bigl(\mathcal{O}_{i,j} - \mathcal{O}^{\theta}_{j}(X_i)\bigr)^2
- \delta \mathcal{O}_{i,j}^2
\right],
\end{equation}
where $\delta\mathcal{O}_{i,j}$ is a per-point tolerance that absorbs known uncertainties 
in the training data. For HRG data, we set $\delta\mathcal{O}_{i,j} = 0$, recovering the 
standard MSE and enforcing exact reproduction of the model. For LQCD data, the tolerance 
is taken as
\begin{equation}
\delta \mathcal{O}_{i,j}
= 0.01 \left(\frac{\mu_{B,i}}{T_i}\right)^4 \mathcal{O}_{i,j},
\end{equation}
reflecting the growth of Taylor truncation errors with $\mu_B/T$. A rigorous estimate of these errors would require higher-order susceptibilities; here, we adopt this simple scaling 
as a practical approach.

Thermodynamic stability requires
\begin{equation}
    \frac{\partial s}{\partial T} > 0, \qquad
    \frac{\partial n_j}{\partial \mu_j} > 0, \quad j \in \{B,Q,S\},
\end{equation}
which must hold not only on the training dataset but throughout the full domain 
$\mathcal{R}_X$. Following the PINN strategy, we enforce these conditions via a penalty 
evaluated at $N_c = N/4$ collocation points sampled uniformly over $\mathcal{R}_X$,
\begin{equation}\label{eq:stab}
\mathcal{L}_{\mathrm{stab}} =
\frac{1}{N_c}\sum_{k=1}^{N_c}\sum_{j}
\max\!\left(0,\,
- \frac{\partial^2 P_{\theta}(X_k)}{\partial X_j^2}
\right).
\end{equation}
Since the stability conditions do not fully determine the EoS there is still some freedom in the extrapolation region. To suppress unphysical oscillations and improve reproducibility across training runs, we add a regularization term penalizing large second derivatives,
\begin{equation}
\mathcal{L}_{\mathrm{reg}} =
\frac{1}{N_c}\sum_{k=1}^{N_c}\sum_{i,j} (-1)^{\delta_{j,0}}
\left(
\frac{\partial^2 P_{\theta}(X_k)}{\partial X_i \partial X_j}
\right)^2,
\end{equation}
evaluated at the same collocation points as Eq.~\eqref{eq:stab}. The sign convention $(-1)^{\delta_{j,0}}$ encourages rapid growth of thermodynamic 
densities with temperature (and vice versa for chemical potentials), allowing a smoother connection to the HRG model in the extrapolation region and suppressing large values of the speed of sound. We note that this choice is phenomenological and can be adjusted depending on the desired behavior of the EoS in the extrapolation domain.

The full objective is
\begin{align}
\mathcal{L}(t) =\,
&\mathcal{L}_P
+ \lambda_1(t)\left(
\mathcal{L}_{\varepsilon}
+\mathcal{L}_{s}
+\mathcal{L}_{n_B}
+\mathcal{L}_{n_Q}
+\mathcal{L}_{n_S}
\right) \nonumber \\
&+ \lambda_2(t)\,\mathcal{L}_{\mathrm{stab}}
+ \lambda_3(t)\,\mathcal{L}_{\mathrm{reg}},
\end{align}
minimized with the Adam optimizer \cite{kingma:2017} over five stages $t = 1,\ldots,5$, with learning rate 
decaying from $\mathrm{lr}_1 = 2^{-10}$ by a factor of $4$ every $200$ epochs down to $\mathrm{lr}_5 = 2^{-18}$. The pressure loss $\mathcal{L}_P$ is active throughout all stages. From $t \ge 2$ onward, once the network has captured the coarse structure of $P_\theta$ on $\mathcal{D}$, the remaining observable losses are activated with $\lambda_1(t \ge 1) = 1$. The stability and regularization weights, which are activated from the third stage,
\begin{equation}
\lambda_2(t \ge 3) = 10^{-4}, \qquad
\lambda_3(t \ge 3) = 5 \times 10^{-9},
\end{equation}
are chosen to keep these terms subdominant relative to the primary data losses, so that they act as mild constraints rather than dominant drivers of the optimization.

In principle, the loss function can include observables that are second-order derivatives of the pressure, such as susceptibilities $\chi^{BQS}_{ijk}$ or the speed of sound $c_s^2$. Their inclusion substantially complicates network training without yielding a significant 
improvement in the resulting EoS. Since the primary application of this work is to heavy-ion collision simulations, where bulk dynamics is governed by lower-order thermodynamic quantities, we do not include these terms in the present construction.

\section{Results}
\subsection{Reproduction of the Hadron Resonance Gas at Low Densities}

Once trained, the EoS can be incorporated directly into hydrodynamic models of heavy-ion collisions. Although the construction is four-dimensional, it is standard practice to restrict it to a two-dimensional subspace by imposing
\begin{align}
    n_Q &= 0.4\, n_B, \nonumber \\
    n_S &= 0, \label{eq:QS_constraints}
\end{align}
reflecting the initial proton-to-baryon ratio  ($Z/A \approx 0.4$ in Au+Au collisions) and the condition of vanishing net strangeness. These constraints allow us to determine $\mu_Q$ and 
$\mu_S$ numerically as functions of $(T, \mu_B)$. In hydrodynamic simulations, $T$ and $\mu_B$ are reconstructed from the local energy density $\varepsilon$ and baryon number 
density $n_B$, with the EoS closing the system of equations.

\begin{figure}[t]
    \centering
    \includegraphics[width=\linewidth]{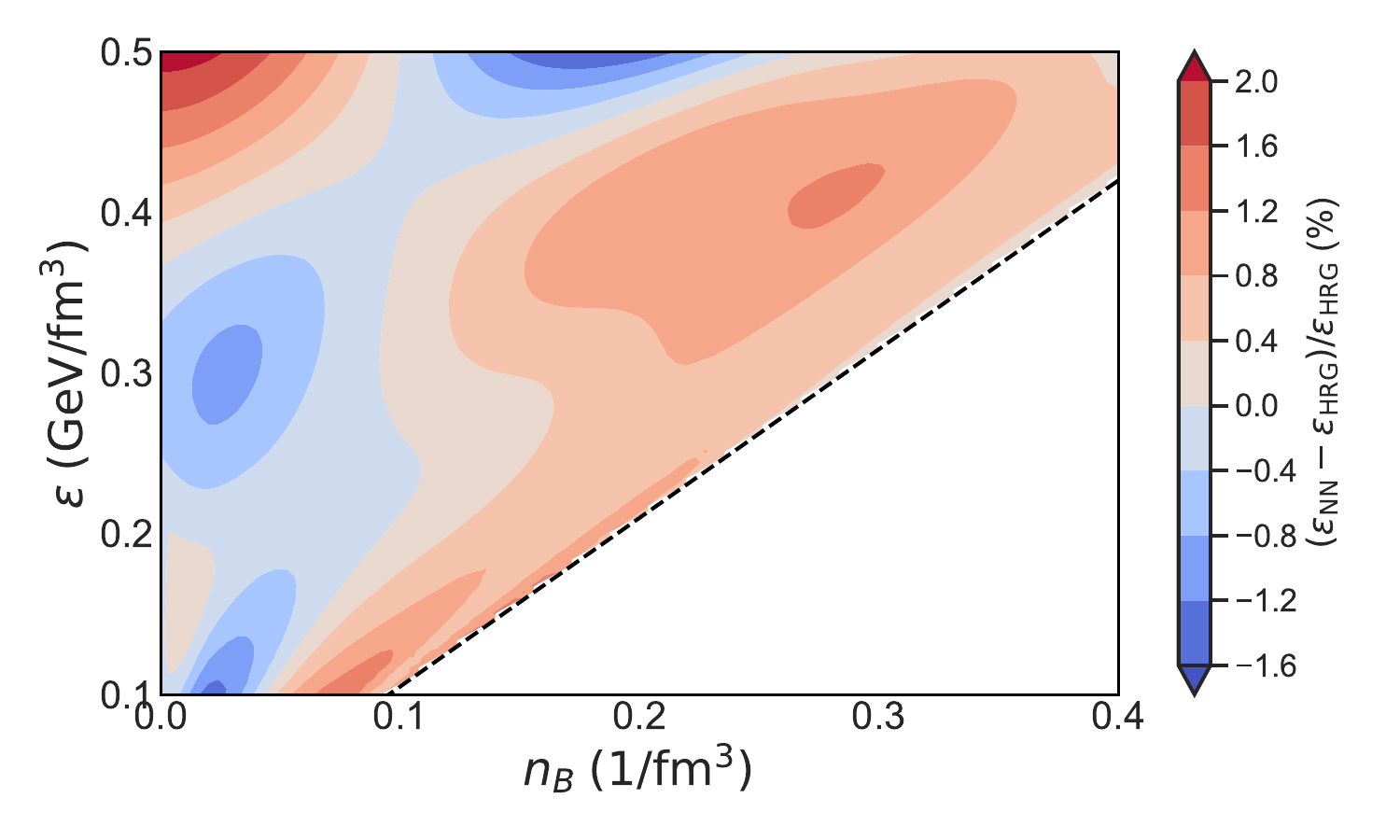}
    \caption{
        Relative difference between the neural-network EoS and the HRG model in the energy 
        density $\varepsilon$, shown in the $(\varepsilon, n_B)$ plane under the constraints 
        $n_Q = 0.4\,n_B$ and $n_S = 0$. The blank region below the black dashed line lies 
        outside the EoS domain $\mathcal{R}_X$ defined in Eq.~(\ref{eq:domain}).
        }
    \label{fig:delta_eps}
\end{figure}

The trained network, together with example code for reconstructing 
$(T, \mu_B, \mu_Q, \mu_S)$ from $(\varepsilon, n_B)$ under the 
constraints~(\ref{eq:QS_constraints}), is publicly available 
at~\cite{Adzhymambetov:2025eos}. To assess the reproducibility of the HRG model, we reconstruct $(T, \mu_B, \mu_Q, \mu_S)$ for $0.1 \leq \varepsilon \leq 0.5~\mathrm{GeV/fm^3}$ and $0 \leq n_B \leq 0.4~\mathrm{fm^{-3}}$ from our model, and then calculate $(\varepsilon, n_B)$ from the HRG model. 
Figure~\ref{fig:delta_eps} shows the relative difference between the neural-network prediction and the HRG calculation. 
The maximum relative error is $2.2\%$, while the mean absolute error over this region is $0.6\%$.

The largest deviation, observed at high $\varepsilon$ and low $n_B$, originates from a slight mismatch 
between the HRG and LQCD equations of state in their overlap region (Fig.~\ref{fig:diagram}). 
Enforcing continuity between the two EoSs leads to a mild underestimation at intermediate $n_B$. 
Although these deviations remain small, improved conservation-law accuracy at particlization can be achieved 
by choosing a switching energy density $\varepsilon_{\mathrm{sw}} \leq 0.45~\mathrm{GeV/fm^3}$.

We note that the neural network does not perform well below $\varepsilon \lesssim 
0.05~\mathrm{GeV/fm^3}$. In practice, this is not a limitation for heavy-ion 
simulations, since local thermodynamic equilibrium is not expected at such low 
densities and the hydrodynamic stage is typically terminated well before this regime 
is reached. If required, the EoS can be smoothly connected to the HRG model at, for example, $\varepsilon = 0.1~\mathrm{GeV/fm^3}$ using a switching function as employed in Ref.~\cite{Monnai:2019hkn}. Similarly, the EoS can be connected to lattice QCD results at $T > 0.5~\mathrm{GeV}$.

\subsection{Adiabatic Trajectories}

\begin{figure}[t]
    \includegraphics[width=\linewidth]{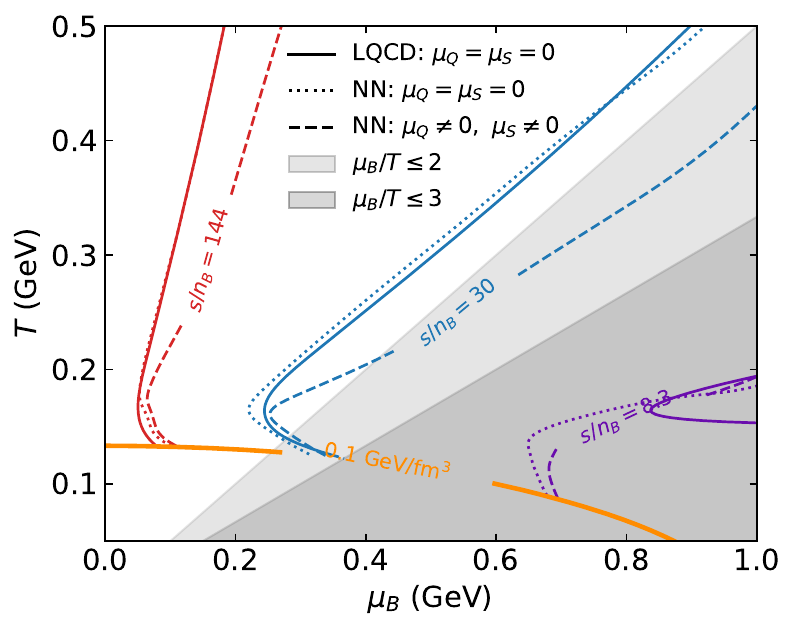}
    \caption{
    Contours of constant $s/n_B$ in the $(T,\mu_B)$ plane. Solid lines: lattice QCD. 
    Dashed and dotted lines: NN predictions with $(\mu_Q,\mu_S)$ defined form constraints~(\ref{eq:QS_constraints}) and with 
    $\mu_Q=\mu_S=0$, respectively. Grey bands mark the regions $\mu_B/T \leq 2$ and 
    $\mu_B/T \leq 3$. The orange contour corresponds to $\varepsilon = 
    0.1~\mathrm{GeV/fm^3}$.
    }
    \label{fig:snB_contours}
\end{figure}

We assess the behavior of the EoS in the extrapolation region by examining adiabatic trajectories, along which the entropy-to-baryon ratio $s/n_B$ is conserved. Figure~\ref{fig:snB_contours} shows contours for $s/n_B = 144$, $30$, and $8.3$, 
commonly associated with Au+Au (or Pb+Pb) collision energies $\sqrt{s_{\mathrm{NN}}} \approx 62.4$, 
$14.5$, and $4.5$~GeV, respectively~\cite{Monnai:2019hkn, Motornenko:2019arp}. Three cases are compared: lattice QCD results at $\mu_Q = \mu_S = 0$, NN predictions at $\mu_Q = \mu_S = 0$, and NN predictions with $\mu_Q$ and $\mu_S$ fixed 
by Eqs.~(\ref{eq:QS_constraints}).

At large $s/n_B$ and $\mu_Q = \mu_S = 0$ the NN reproduces the lattice results nearly exactly above the crossover ($T \gtrsim 0.165~\mathrm{GeV}$), with only minor deviations at lower temperatures, where the model transitions to the HRG description as constructed. At intermediate $s/n_B$, the overall agreement persists, though deviations become visible. At the smallest value, which lies beyond the reach of current lattice calculations, the trajectories exhibit qualitatively different behavior, as 
expected. Imposing the constraints~(\ref{eq:QS_constraints}) noticeably modifies the 
trajectories at high temperatures, consistent with the lattice findings of 
Ref.~\cite{Noronha-Hostler:2019ayj}. Across all cases, the neural network produces smooth 
and physically reasonable trajectories even in the extrapolation region, including a 
turning point near the transition to the HRG regime at low energy densities.

\subsection{Speed of Sound}

\begin{figure}[t]
    \centering
    \includegraphics[width=\linewidth]{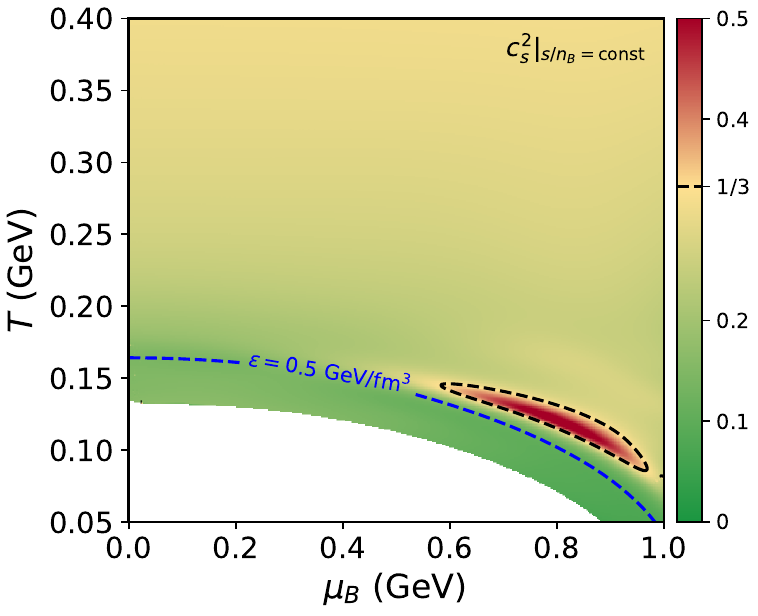}
    \caption{
    Speed of sound squared $c_s^2$ along adiabatic trajectories ($s/n_B = \mathrm{const}$) 
    under the constraints $n_Q = 0.4\,n_B$ and $n_S = 0$. The black dashed line marks the 
    boundary where $c_s^2 = 1/3$.
    }
    \label{fig:c2s}
\end{figure}

The squared speed of sound along adiabatic trajectories is computed as
\begin{equation}
    c_s^2 = \left.\frac{dP}{d\varepsilon}\right|_{s/n_B},
\end{equation}
under the constraints~(\ref{eq:QS_constraints}). The result is shown in 
Fig.~\ref{fig:c2s}. At low $\mu_B$, the familiar dip in the crossover region is 
reproduced, consistent with Ref.~\cite{Noronha-Hostler:2019ayj}. This feature persists 
toward larger $\mu_B$ approximately along the $\varepsilon = 0.5~\mathrm{GeV/fm^3}$ contour, near the transition to HRG.

At large $\mu_B$, the EoS exhibits a region where $c_s^2$ exceeds the conformal limit 
$1/3$, reaching values as high as $c_s^2 \approx 0.5$. This is not unexpected: violations 
of the conformal bound at high baryon density have been extensively discussed in the 
context of dense QCD matter and compact-star physics~\cite{Bedaque:2014sqa, Tews:2018kmu}. 
In particular, neutron-star mass-radius measurements strongly suggest that $c_s^2 > 1/3$ 
in neutron-star cores is required to support two-solar-mass stars~\cite{Annala:2019puf, 
Bedaque:2014sqa}, and analogous behavior has been discussed in heavy-ion collisions at high 
baryon density~\cite{Gong:2024lhq, Motornenko:2019arp}.

The enhancement of $c_s^2$ can be attributed to the rapid growth of entropy in the HRG 
model at high particle densities, where excluded-volume corrections and interaction effects 
are expected to become significant. Notably, this region — separated by the black dashed 
line in Fig.~\ref{fig:c2s} — can be eliminated by shifting the HRG transition to slightly 
lower temperatures at large $\mu_B$. Specifically, lowering the transition condition from 
$\varepsilon = 0.5$ to $0.4~\mathrm{GeV/fm^3}$ (a threshold that could in principle depend 
on $\mu_B$) keeps $c_s^2$ below $1/3$ throughout. The presence of a phase transition, not 
included in the current construction, must also qualitatively modify this picture. Since 
particlization at constant energy density is the most widely used prescription in hybrid 
models, and the hadronic afterburner is well described by transport in the hadronic cascade 
regime, we do not incorporate these modifications in the present study.

\subsection{Application in a Hybrid Model}

\begin{figure}[t]
    \centering
    \includegraphics[width=\linewidth]{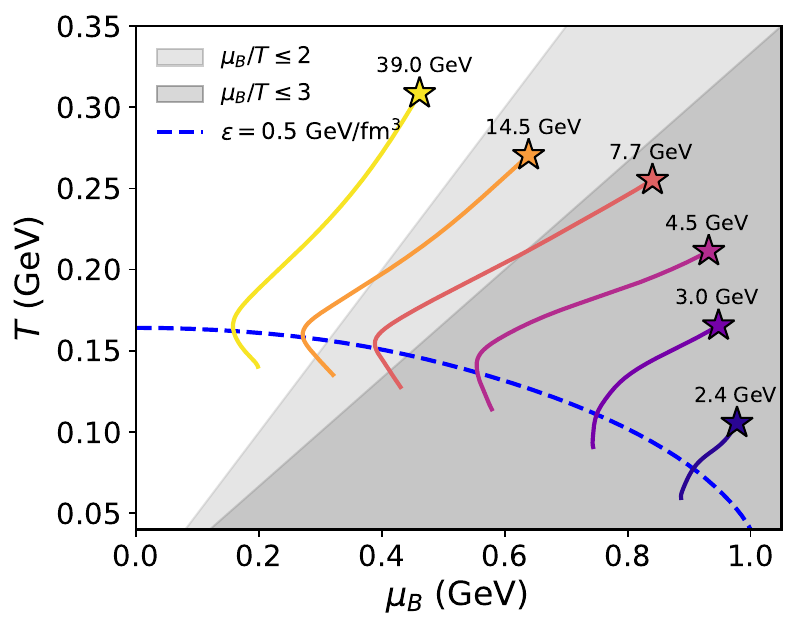}
    \caption{
    Adiabatic trajectories of constant $s/n_B$ for different collision energies in iHKM 
    simulations of $0$--$10\%$ central Au+Au collisions. Each trajectory passes through 
    the median switching point $(T, \mu_{B}, \mu_Q, \mu_S)_{\mathrm{sw}}$ on the 
    particlization hypersurface and extends from the maximal initial energy density 
    $\varepsilon_{\max}$ down to $\varepsilon = 0.1~\mathrm{GeV/fm^3}$. Markers indicate 
    the median switching points; labels show $\sqrt{s_{\mathrm{NN}}}$.
    }
    \label{fig:ihkm}
\end{figure}

As a demonstration, we implement the EoS in the integrated hydrokinetic model 
(iHKM)~\cite{Adzhymambetov:2024zzz} and simulate $0$--$10\%$ central Au+Au collisions at several energies. Initial conditions are generated by UrQMD~\cite{Bleicher:1999xi} in 
cascade mode, followed by a gradual transition to viscous hydrodynamics and a hadronic 
afterburner again described by UrQMD. The constraints~(\ref{eq:QS_constraints}) are 
enforced throughout the hydrodynamic stage.

We do not optimize model parameters, but instead adopt the parametrization of Ref.~\cite{Rathod:2025gvj} extrapolated to lower energies, including the hydrodynamic initialization time, called thermalization time in iHKM
\begin{equation}
\tau_{\mathrm{th}} = \tau_0 + \tau_{\mathrm{rel}}
= 0.75\,\tau_{\mathrm{overlap}}(\sqrt{s_{\mathrm{NN}}}) + 1.0~\mathrm{fm}/c,
\end{equation}
where $\tau_{\mathrm{overlap}}$ is the nuclear crossing time assuming free streaming. Initial energy and conserved-charge densities are smeared with a Gaussian of width 
$R = 1.0~\mathrm{fm}$, and particlization is performed at the constant switching energy 
density $\varepsilon_{\mathrm{sw}} = 0.5~\mathrm{GeV/fm^3}$.

From each simulation from the constructed particlization hypersurface, we extract the median (sorted by $\mu_B$) switching point $X_{\text{sw}}=(T, \mu_B, \mu_Q, \mu_S)_{\mathrm{sw}}$, and from hydrodynamic stage and the maximal energy density $\varepsilon_{\max}$ at $\tau_{\text{th}}$. Figure~\ref{fig:ihkm} shows the adiabatic trajectories passing through $X_{\text{sw}}$, extending from $\varepsilon_{\max}$ down to $\varepsilon = 0.2~\mathrm{GeV/fm^3}$. We note that hydrodynamic simulations involve substantial spatial inhomogeneities in the initial state, so the actual thermodynamic trajectories of individual fluid cells can deviate significantly from these representative paths, with different regions of the fireball tracing distinct routes through the diagram.

Figure~\ref{fig:ihkm} nonetheless illustrates the key motivation for the present work. Already at $\sqrt{s_{\mathrm{NN}}} = 7.7~\mathrm{GeV}$, the lowest energy of the RHIC BES program~\cite{STAR:2010vob}, the system accesses regions with $\mu_B/T \gtrsim 3$, beyond the reach of lattice QCD. At lower energies, such as those explored at HADES~\cite{HADES:2009aat} and FAIR/CBM~\cite{CBM:2016kpk}, the fireball may probe baryon chemical potentials up to $\mu_B \sim 1~\mathrm{GeV}$, a regime where the EoS must rely on effective descriptions such as the one presented here.

\section{Summary and Outlook}

We have constructed a four-dimensional equation of state for strongly interacting matter at finite temperature and conserved charge chemical potentials, designed for use in hybrid models of relativistic heavy-ion collisions. The construction employs a deep neural network trained on two complementary inputs: lattice QCD results via a Taylor expansion in chemical potentials based on work \cite{Noronha-Hostler:2019ayj}, and hadron resonance gas thermodynamics at low temperatures and densities.

A central requirement of the construction is precise matching to the HRG equation of state below $\varepsilon = 0.5~\mathrm{GeV/fm^3}$, the typical switching criterion used for particlization in hybrid simulations. This ensures a smooth transition between the hydrodynamic and hadronic cascade stages, with no significant discontinuities in either the thermodynamic 
variables $(T, \mu_B, \mu_Q, \mu_S)$ nor in the derived bulk quantities 
$(\varepsilon, n_B, n_Q, n_S, P)$. Thermodynamic consistency across the full domain is enforced via physics-informed loss terms that penalize violations of stability conditions at collocation points sampled throughout the extrapolation region.

The equation of state covers a wide region of the QCD phase diagram, with $\mu_B$ up to $1.0~\mathrm{GeV}$, making it applicable to hybrid-model simulations at collision energies relevant to the RHIC Beam Energy Scan, HADES, and FAIR/CBM. We have demonstrated its use in iHKM simulations of central Au+Au collisions at several energies, showing that 
already at $\sqrt{s_\mathrm{NN}} = 7.7~\mathrm{GeV}$ the system probes baryon chemical potentials beyond the reach of direct lattice QCD calculations, and that at lower energies values of $\mu_B \sim 1~\mathrm{GeV}$ can be accessed.

By construction, the present equation of state does not include a phase transition or critical endpoint. Incorporating such structures, for example, by embedding a critical point, is a natural direction for future work. Further extensions include systematic optimization of the switching conditions as a function of $\mu_B$, the inclusion of second-order derivative thermodynamics in the loss function, and modifications to the HRG model, such as excluded volume or mean-field interactions.

The trained neural network weights, code for EoS table construction, and ready-to-use example tables are publicly available at~\cite{Adzhymambetov:2025eos}.

\begin{acknowledgments}
This work was supported by the National Research Foundation of Ukraine, grant No.~2025.07/0050.
\end{acknowledgments}

\bibliographystyle{apsrev4-2} 
\bibliography{apssamp}

\end{document}